\documentclass[runningheads]{llncs}
\usepackage{mathtools}
\usepackage{graphicx}

\newcommand{\kij}{k_{ij}}

\newcommand{\aij}{\alpha_{ij}}
\begin{document}

\title{Modeling and inference of spatio-temporal protein dynamics across brain networks}
\titlerunning{Protein dynamics brain network}
\author{Sara Garbarino\inst{} \and Marco Lorenzi\inst{} \\
for the Alzheimer's Disease Neuroimaging Initiative \thanks{Data used in preparation of this article were obtained from the Alzheimer’s Disease Neuroimaging Initiative (ADNI) database (adni.loni.usc.edu). As such, the investigators within the ADNI contributed to the design and implementation of ADNI and/or provided data but did not participate in analysis or writing of this report. }
}
\institute{EPIONE project-team, INRIA, Universite Cote d'Azur,  Sophia Antipolis, France
\email{sara.garbarino@inria.fr}}

\maketitle
\begin{abstract}

Models of misfolded proteins (MP) aim at discovering the bio-mechanical propagation properties of neurological diseases (ND) by identifying plausible associated dynamical systems. Solving these systems along the full disease trajectory is usually challenging, due to the lack of a well defined time axis for the pathology. This issue is addressed by disease progression models (DPM) where long-term progression trajectories are estimated via time reparametrization of individual observations. However, due to their loose assumptions on the dynamics, DPM do not provide insights on the bio-mechanical properties of MP propagation.
Here we propose a unified model of spatio-temporal protein dynamics based on the joint estimation of long-term MP dynamics and time reparameterization of individuals observations. The model is expressed within a Gaussian Process (GP) regression setting, where constraints on the MP dynamics are imposed through non--linear dynamical systems. We use stochastic variational inference on both GP and dynamical system parameters for scalable inference and uncertainty quantification of the trajectories. Experiments on simulated data show that our model accurately recovers prescribed rates along graph dynamics and precisely reconstructs the underlying progression. When applied to brain imaging data our model allows the bio-mechanical interpretation of amyloid deposition in Alzheimer's disease, leading to plausible simulations of MP propagation, and achieving accurate predictions of individual MP deposition in unseen data.

\end{abstract}

\section{Introduction}

A peculiarity of neurodegenerative diseases (NDs) is the misfolding and subsequent accumulation of pathological proteins in the brain, leading to cellular dysfunction, loss of synaptic connections, and neuronal loss \cite{soto2012protein}. Misfolded protein (MP) aggregates can self-propagate and spread the pathology between cells and tissues, along functional or structural brain networks \cite{junker2013self}. 

With the aim of describing such processes, a variety of mathematical models has been proposed for providing better insight into the microscopic kinetic dynamics governing the processes of proteins propagation \cite{carbonell2018mathematical}. 
Complementarily, another class of models relies on macroscopic measurements from molecular and structural imaging for describing the effects of MP propagation along brain networks \cite{raj2012network,oxtoby2017datadriven,zhou2012predicting,iturria2018multifactorial,cauda2018brain}. 
 Such MP kinetics models are of strategic relevance, as they may provide new understanding of the mechanisms involved in NDs, and thus allow identification of novel strategies for treatment and diagnosis. 
These models usually define the propagation dynamics through diffusion equations. This modelling choice allows to reduce the number of parameters to be estimated, but comes at the expenses of an oversimplification of the dynamics governing the MP process.
First, while the pathological kinetics may be assimilated to diffusive processes in short term observations, the long term evolution of NDs are unlikely to have diffusive properties. For example, the asymptotically constant behaviour of NDs may not be described by the stationary and constant rate of change specified by diffusion equations. Second, such models require a precise definition of the time axis, which is typically not well defined in clinical data sets. To address this issue, several alternative disease progression models (DPM) have been proposed \cite{young2014datadriven,lorenzi2017probabilistic,schiratti2017bayesian,donohue2014estimating}.
These approaches allow to reconstruct biomarkers trajectories along the long term disease progression by ``stitching'' together short term individual measurements.
Each subject is characterized by specific time parameters quantifying their pathological stage with respect to the estimated global group--wise evolution. However, these models provide an ``apparent'' description of biomarkers dynamics, without in fact elucidating the kinetics and relationships across biomarkers. 

To date, no modelling framework allows for joint MP kinetics modelling and reconstruction of the biomarkers dynamics across the whole disease long term evolution. The problem is challenging since it requires to simultaneously account for short term observations to reconstruct the long term disease progression, and to estimate group--wise dynamics parameters specified by high--dimensional dynamical systems.

In this paper we solve this problem by formulating a model for the dynamics of MP accumulation, clearance and propagation (ACP) across structural brain networks, which includes data--driven estimates of the long term protein trajectories from short term data. Figure 1 shows a schematic representation of our framework.
The ACP model is formulated as a constrained regression problem in a Bayesian non--parametric setting, where the MP progression is modelled by a Gaussian Process (GP), and constraints on the MP dynamics are imposed through systems of non linear ordinary differential equations (ODEs). The Bayesian setting allows for uncertainty quantification of the MP dynamics while, to achieve tractability, the inference problem is solved via stochastic variational inference.
 \begin{figure}[ht]\label{fig:pictorial2}
    \centering
    \includegraphics[width=4.2in]{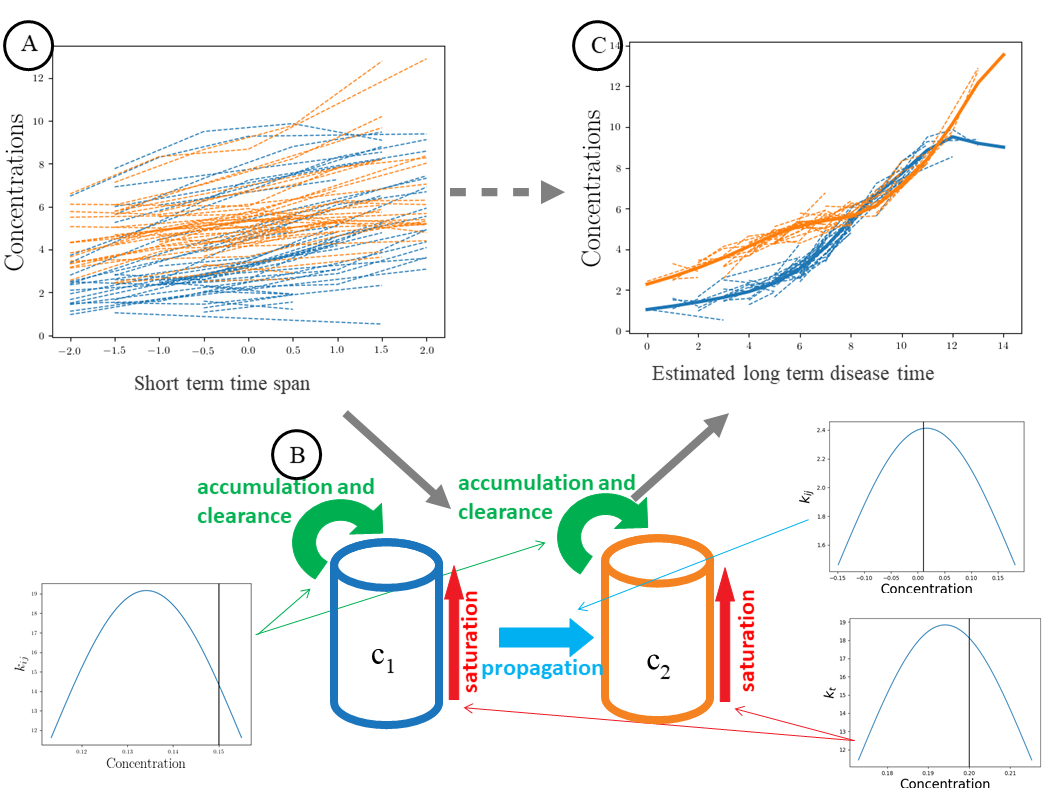}
    \caption{Schematic representation of our framework. Here we have two brain regions whose MP concentrations $c_1$ and $c_2$ are collected for many subjects over a short term time span (A). The dynamics of such concentrations is described in terms of the accumulation, clearance and propagation processes, with unknown parameters (B). The proposed Bayesian framework estimates the distribution of such parameters (here plotted against their ground truth values - the vertical black lines in the distributions subplot), and the long term trajectories with respect to the estimated disease time axis (C).}
\end{figure}
The constrained regression framework provides a complete description of the MP dynamics, which can be subsequently used for simulating and predicting MP changes over time through forward integration of the estimated dynamical systems. The estimated MP dynamics also provide an instrument to investigate different hypotheses of MP propagation.

We test our framework against synthetic data and compare its performances in recovering the simulated evolution and the time reparameterization as compared to standard disease progression models based on monotonic constraints.
Finally, we demonstrate our framework on AV45-PET data of Alzheimer's Disease (AD) subjects from the ADNI data set. 
We show that it allows to compare different hypothesis of MP kinetics: diffusive vs non--linear and time--varying dynamics properties (ACP). We show that the ACP model outperforms diffusive ones in terms of prediction of amyloid deposition in unseen follow-up data.

\section{Methods}

\subsection{Non--linear and time--varying MP kinetics model}
We consider the brain as a system of $N$ interconnected regions, where each region $i$ ($i=1,...,N$) is characterized by its concentration $c_i(t)$ of MP proteins along time.
Standard MP kinetics models are based on the definition of dynamical systems of the form
\begin{equation}\label{eq:stateart}
    \dot{\boldsymbol{c}}(t) = \beta A\boldsymbol{c}(t),
\end{equation}
where $\boldsymbol{c}(t)$ is the vector of concentrations of MP across brain regions, $A$ is a diffusion matrix and $\beta$ is the parameter describing MP propagation. The operator $A$ is usually defined as the graph Laplacian of the brain connectome, while $\beta$ is typically assumed to be constant throughout the whole disease progression \cite{raj2012network}.

Here we introduce an extension of this paradigm which accounts for the dynamics characterized by the time--varying and non--linear parameters of MP accumulation, brain response via MP clearance, and long term propagation across neighbouring neuronal cells: the ACP model. Within this setting, equation (\ref{eq:stateart}) is reformulated as
\begin{equation}\label{eq:mainmatr}
 \dot{\bf{c}}(t) = M_{ACP}({\bf c}(t)) {\bf c}(t),
\end{equation}
where the matrix $M_{ACP}$ is factorized into (dependence on $t$ is omitted but implied):
 $M_{ACP}(\boldsymbol{c}) = M_{AC}(\boldsymbol{c}) - M_{out}(\boldsymbol{c}) + M_{in}(\boldsymbol{c})$.                
Here, $M_{AC}(\boldsymbol{c})$ accounts for the total aggregation of MP plaques, i.e. sum of accumulation and clearance, while the remaining two matrices describe long--range propagation from ($M_{out}(\boldsymbol{c})$) or to ($M_{in}(\boldsymbol{c})$) for each region.
Our assumptions on the MP dynamics are the following:
 \begin{itemize}
    \item no aggregation nor propagation occur in healthy conditions. MP plaques aggregation develops when the accumulation-clearance equilibrium breaks. This can be modelled with the assumption that $\bar{k}_a - \bar{k}_c > 0 $, where  $\bar{k}_{\cdot}$ are the maximum rates of accumulation and clearance and are assumed to be constant across regions. We define $\bar{k}_t := \bar{k}_a - \bar{k}_c$ as the maximum rate of total aggregation.
    
    \item We hypothesize a region--dependent critical threshold $\eta_i$ above which the aggregation process reaches a plateau. This is modelled by a sigmoid function 
         \begin{equation}\label{eq:kt}
            \displaystyle  k_t(c_i) = \frac{\bar{k_t}}{1+e^{l_2(c_i-\eta_i)}}.
        \end{equation}\textbf{}
    
    \item When passing a critical threshold $\gamma_j$ the MP concentration in each region $j$ saturates and triggers propagation towards the connected regions. Also, it reaches a plateau when passing a threshold $\eta_j$. Again this can be modelled by a function asymptotically dropping to zero:
        \begin{equation}\label{eq:kij}
            \displaystyle \kij(c_j) = \frac{\bar{k}_{ij}}{\left(1+e^{-l_1(c_j-\gamma_j)}\right)\left(1+e^{l_2(c_j-\eta_j)}\right)},
        \end{equation}
        representing the non--linear rate of propagation from region $j$ to region $i$. Here $\bar{k}_{ij}$ is the maximum rate of propagation between the two regions, and we assume $\bar{k}_{ij}=\bar{k}_{ji}$. We combine the propagation coefficients in a matrix describing the global brain--scale propagation process: $K(\boldsymbol{c}) = \left(\kij(c_j)\right)_{ij}$.
        
    \item The substrate for propagation is the structural connectome, here represented by the symmetric and normalized adjacency matrix of connections between brain regions $A=(\aij)$.
 \end{itemize}
 Such assumptions are formalized into the following functionals:
 \begin{equation}\label{eq:mat1}
\left(M_{AC}({\bf c})\right)_{ij} = 
\begin{cases}
k_t({\bf c})  \quad \text{if} \quad  i=j \\ 
0\quad  \text{otherwise};
\end{cases}
 \end{equation}
  \begin{equation}\label{eq:mat2}
M_{in}({\bf c}) = K({\bf c})\odot A;
 \end{equation}
   \begin{equation}\label{eq:mat3}
\left(M_{out}({\bf c})\right)_{ij} = \begin{cases}
\sum_{j}(K({\bf c})_{ij} \odot A_{ij}) \quad \text{if} \quad  i=j \\ 
0\quad  \text{otherwise}.
\end{cases}
 \end{equation}
Overall, the ACP model depends on $1 + 2N +\frac{N^2-N}{2}$ parameters:  ${\bf \theta} =(\bar{k_t}, \eta_i, \gamma_i , \bar{\kij})$ for $i,j=1,...,N$.

\subsection{Extending MP dynamics modelling to account for time reparametrization}
Once defined our dynamical system as the one described in (\ref{eq:mainmatr}), we need to incorporate it within a regression framework for short term data.
Let us assume to have $S$ subjects for which we have measures of MP concentrations ${\bf C}$ in $N$ brain regions at different time--points, encoded in a vector ${\bf t}$: ${\bf C}$ is therefore the realization of $\boldsymbol{c}(t)$ at times $\boldsymbol{t}$. For notation simplicity we assume here ${\bf t}$ to be the same for every subject, but computations extend easily to more general cases. 

The observations ${\bf C}$ for subject $k$ at a time points ${\bf t}$ can be modelled as a random sample from the following generative model \cite{lorenzi2017probabilistic}:
\begin{equation}\label{eq:generative}
 {\bf C}^k(\boldsymbol{t}) = {\bf f}(\boldsymbol\tau^k(\boldsymbol{t})) + {\boldsymbol\nu}^k +{\boldsymbol\epsilon}.
\end{equation}
Here ${\bf f}$  is the fixed effect function modeling the concentrations' longitudinal evolution and is modelled as a GP;
${\boldsymbol \tau}^k(\boldsymbol{t})$ is the individual time reparametrization with respect to the global group--wise evolution, and is modelled as a linear shift ${\boldsymbol \tau}^k = {\bf t}^k + d^k$;
$\boldsymbol\nu^k$ is the individual random effect, assumed to be Gaussian correlated perturbations $\mathcal{N}(0,\phi_N^k)$; $\boldsymbol\epsilon$ is the observational noise. 
We introduce constraints on the dynamics of the model $\boldsymbol{f}$ enforcing the concentrations' evolution to the ACP model. This means specifying a family of admissible functions whose derivatives evaluated at the inputs $\boldsymbol{t}$ satisfy the ACP constraint:
\begin{equation}\label{eq:const}
\mathcal{A} = \{\boldsymbol{f}(t) : \boldsymbol{\dot{f}}(\boldsymbol{t}) = M_{ACP}(\boldsymbol{f}(\boldsymbol{t}))\boldsymbol{f}(\boldsymbol{t})\}.
\end{equation}
We note that the constraints are imposed only on the group--wise dynamics $\boldsymbol{f}$ and not on the random--effects. This is done to reduce complexity and the model's parameters. Relaxing the constraints at individual level is also meaningful, as some subjects may be characterized by potentially different dynamics due to specific clinical conditions.

\subsection{The inference scheme}

We define as $\boldsymbol{F}^k$  the realization of $\boldsymbol{f}$ at $\boldsymbol\tau^k(\boldsymbol{t})$, and as $\boldsymbol{\dot{F}}^k$ the set of realizations of $\boldsymbol{f}$ and of its derivatives at $\boldsymbol\tau^k(\boldsymbol{t})$. 
We also indicate by $\boldsymbol{F}$, $\boldsymbol{\nu}$, and $\boldsymbol{\dot{F}}$ the collections of $\boldsymbol{F}^k$,  $\boldsymbol{\nu}^k$, and $\boldsymbol{\dot{F}}^k$ for all the subjects ($k=1,...,S$).
Similarly, we define $\boldsymbol{\tau}$ as the collections of $\boldsymbol{\tau}^k$. We denote by $\theta$ the set of parameters for the MP dynamics, and by $\phi_N$ the parameters associated to $\boldsymbol{\nu}$.
Our framework is formulated as the constrained regression defined through two likelihood elements: 
a data fidelity term $p(\boldsymbol{C}|\boldsymbol{F},\phi_N, \boldsymbol{\tau},\epsilon)$ and a constraint term $p(\mathcal{A}|\boldsymbol{\dot{F}}, \theta, \boldsymbol{\tau},\zeta)$, where $\epsilon$ and $\zeta$ are the associated noise parameters. 

Following \cite{lorenzi2018constraining} we solve the constrained regression problem by determining a lower bound for the marginal function 
\begin{equation}\label{eq:marginal}
\begin{split}
    p(\boldsymbol{C},\mathcal{A}|\phi_N,\boldsymbol{\tau},\boldsymbol{t},    \epsilon,\zeta) = 
\int &p(\boldsymbol{C}|\boldsymbol{F},\phi_N,\boldsymbol{\tau},\boldsymbol{t},\epsilon)p(\mathcal{A}|\boldsymbol{\dot{F}},\theta,\boldsymbol{\tau},\boldsymbol{t},\zeta) \\
&p(\boldsymbol{F},\boldsymbol{\dot F}|\phi_N,\boldsymbol{\tau},\boldsymbol{t})p(\theta)d\boldsymbol{F}d\boldsymbol{\dot F}d\theta,
\end{split}
\end{equation}
where 
\begin{equation}
    p(\boldsymbol{F},\boldsymbol{\dot F}|\phi_N,\boldsymbol{\tau},\boldsymbol{t})d\boldsymbol{F} = p(\boldsymbol{\dot F}|\boldsymbol{F})p(\boldsymbol{F}|\phi_N,\boldsymbol{\tau},\boldsymbol{t}).
\end{equation}
We assume the likelihood for data and constraints to be respectively Gaussian and Student--t with  parameters $\epsilon$ and $\zeta$ \cite{lorenzi2018constraining}, and we approximate the GP via random features expansion, as shown in \cite{cutajar2017random}. 
Specifically, the GP realizations can be expressed as $\boldsymbol{F}\approx h(\boldsymbol{t}\boldsymbol{\Omega})\boldsymbol{W}$, where $\boldsymbol{\Omega}$ is a linear projection of the input $\boldsymbol{t}$ into the random feature space specified by the trigonometric activation functions $h(\cdot)=(\cos(\cdot),\sin(\cdot))$, and $\boldsymbol{W}$ are the regression parameters. Such approximation extends to the derivatives of the GP thanks to the chain rule  \cite{lorenzi2018constraining}. As a result, the GP function and its derivatives can be both identified by the parameters $\boldsymbol{W}$ and $\boldsymbol{\Omega}$.

Solving (\ref{eq:marginal}) amounts at doing inference on $\boldsymbol{F}$, which in this setting means inference on $\boldsymbol{W}$ and $\boldsymbol{\Omega}$.
Following \cite{cutajar2017random}, we optimize (\ref{eq:marginal}) through variational inference of $\boldsymbol{W},\boldsymbol{\Omega}$ and $\theta$. This leads to the optimization of the evidence lower bound (ELBO):

\begin{equation}\label{eq:elbo}
\begin{split}
    \log(p(\boldsymbol{C},\mathcal{A}|\phi_N,\boldsymbol{\tau},\boldsymbol{t},    \epsilon,\zeta)) \geq& E_{q(\boldsymbol{W})}\left[\log(p(\boldsymbol{C}|\boldsymbol{\Omega}, \boldsymbol{W}, \phi_N,\boldsymbol{\tau},\boldsymbol{t},\epsilon))\right]+\\
    &+ E_{q(\boldsymbol{W})q(\theta)}\left[\log(p(\mathcal{A}|\boldsymbol{\Omega}, \boldsymbol{W}, \theta,\boldsymbol{\tau},\boldsymbol{t},\zeta))\right]+\\
    &- DKL(q(\boldsymbol{W})|p(\boldsymbol{W})) - DKL(q(\theta)|p(\theta)).
\end{split}
\end{equation}
Here $DKL(q|p)$ is the Kullback Leibler divergence between $p$ and its variational approximation $q$; we assume $q(\boldsymbol{W})$ and $q(\theta)$ to be Gaussian.
Details on the implementation setting are in Supplementary Material.

\section{Simulation Results}
We test the ability of our framework in reconstructing the long term trajectories of the ACP dynamical system from noisy samples of short term data (Figure 2). Results are compared to the ones obtained by using the GP Progression Model \cite{lorenzi2017probabilistic}, which includes a monotonicity constraints on the trajectories. We also test the model with data generated from a single subject and with known time--axis (Supp. Mat.). Synthetic data are generated according to the parameters specified in Table 1.
Figure 2B)-top shows the reconstructed MP trajectories from short-term data in 2A), for a two-dimensional test set.
\begin{table}[htb!]
    \centering
    \begin{tabular}{c|c|c|c|c}
         N subjects & N regions & time interval & time--points per subject & noise \\
         \hline
         50 & $\{2, 3,11,42\}$ & $[0,15]$ & $\{1,2,3,4\}$& $\mathcal{N}(0,\sigma)$, $0.2\leq\sigma\leq0.4$  \\
    \end{tabular}
    \vspace{0.1cm}
    \vspace{.1cm}
    \begin{tabular}{c|c|c|c}
         $\bar{k}_{ij}$ & $\bar{k}_t$ & $\gamma$ & $\eta$\\
         \hline
         $\mathcal{U}(0,1)$& $\mathcal{U}(0,1/2)$&$\mathcal{U}(1,\max(\frac{\boldsymbol{C}}{2}))$&$\mathcal{U}(\max(\frac{\boldsymbol{C}}{2}),\max(\boldsymbol{C}))$
    \end{tabular}
    \vspace{0.1cm}
    \caption{Synthetic data generation parameters. 
    }
    \label{tab:synthetic}
\end{table}
\begin{figure}[hbt!]\label{fig:2}
    \centering
    \includegraphics[width=5in]{{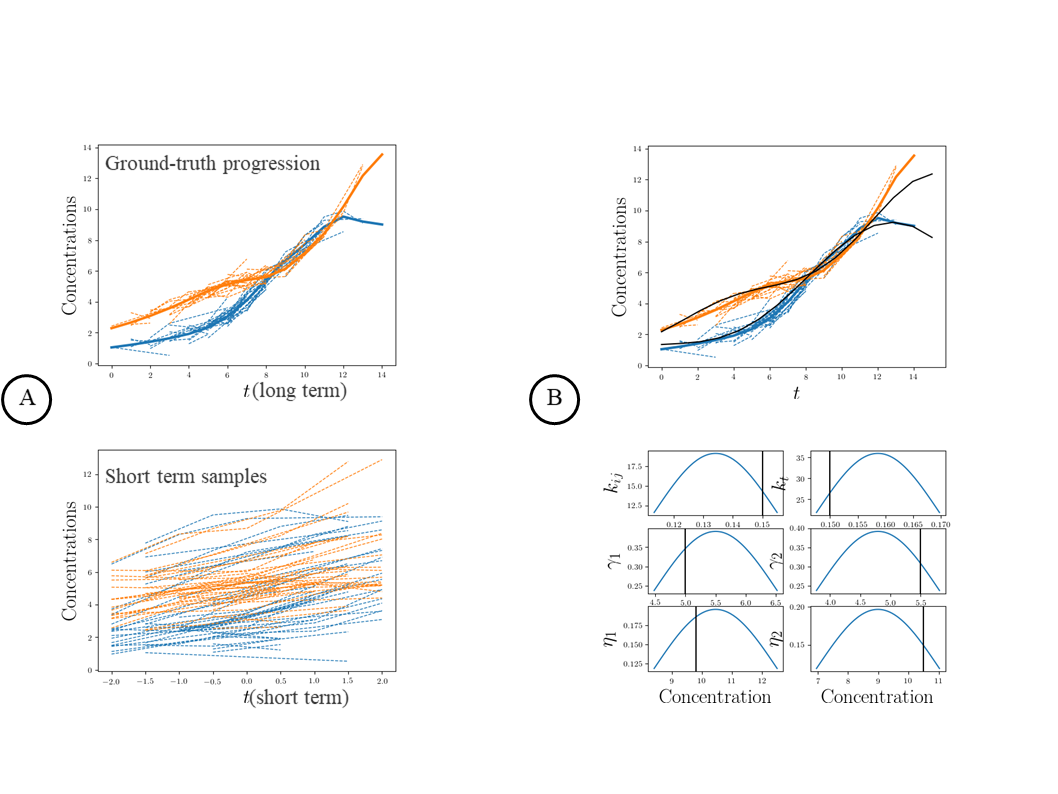}}
    \caption{Results for a 2D example data. A): ground truth GP progressions and associated short term data used for benchmarking. B): ground truth and reconstructed (average) long term trajectories, and 
    reconstructed MP parameters distributions, whose ground truth values are indicated by vertical bars.}
\end{figure}
We run synthetic tests varying the initial values of the MP parameters, the noise and the number of regions. Then, we compared our estimates of the GP and time-shift parameters with results obtained using the GP Progression Model in \cite{lorenzi2017probabilistic}. Table 2 shows results in terms of distributions of root mean squared errors (RMSE), for increasing number of regions. Distributions of RMSE were obtained by sampling 200 times from the estimated distributions. The ACP model generally provides better estimates for the reconstruction of the long term trajectories, as well as for the estimation of the individual time--shift as compared to the standard DPM provided by the monotonic GP. Moreover, while our framework allows the identification of the prescribed dynamical system parameters with high degree of accuracy (Table 2, last row), the monotonic DPM does not allow the estimation of these quantities.
\begin{table}[htb!]
    \centering
    \begin{tabular}{cccccc}
         &&N=2 & N=3 &  N=11 & N=42 \\
        \hline
        \hline
        && & Data fit & & \\
        \hline
        RMSE&ACP GP & 1.17(0.77) & 0.93(0.37) & 1.52(0.25) & 1.07(0.64)\\
        &monotonic GP & 1.32(0.68) & 1.08(0.52) & 1.60(0.29) & 1.20(0.72)\\
        \hline
        && & Time--shift & & \\
        \hline
        RMSE&ACP time-shift & 1.67(0.49) & 1.92(0.64) & 1.53(0.42) & 1.19(0.39)\\
        &monotonic time--shift & 1.87(0.44) & 1.97(0.58) & 1.62(0.42) & 1.20(0.41)\\
        \hline
        && & Dynamical parameters & & \\
        \hline
        relative&ACP GP & 6.3\%&9.6\%& 11.4\%& 21.9\% \\
        error&monotonic GP &--&--&--&--\\
    \end{tabular}
    \caption{RMSE results for GP fit, time--shifts estimates and dynamical parameters for both the GP Progression Model with monotonicity (GP) and the ACP model. The error for the dynamical parameters is expressed, in percentage, relatively to the ground truth parameters.}
    \label{tab:RMSE_ACPdiff}
\end{table}
\vspace{-1.2cm}

\section{Modeling amyloid deposition from imaging data}
\subsection{Data Acquisition and Preprocessing.}
\subsubsection{ADNI data}
This study used 1091 individual data from ADNI, with a total of 2380 longitudinal data. We collected clinical, demographic and AV45-PET SUVr data. All the subjects with either ``Dementia'', ``Mild Cognitive Impairment'' or ``Cognitively Normal'' clinical diagnosis were selected.
The ADNI was launched in 2003 as a public-private partnership, led by Principal Investigator Michael W. Weiner, MD. The primary goal of ADNI has been to test whether serial Magnetic Resonance Imaging (MRI), Positron Emission Tomography (PET), other biological markers, and clinical and neuropsychological assessment can be combined to measure the progression of MCI and early AD. For up-to-date information, see www.adni-info.org.
ADNI AV45-PET SUVR data are already computed on Freesurfer--defined regions, and normalized against the reference region (cerebellum); for full information please see: https://adni.bitbucket.io/reference/docs/. We constructed a regression model for each region separately with normalized SUVr as dependent variable and gender, age, APEO4 genotype and education as independent variables. The residual of the fit is the new value associate to each region. This was done for both patients (AD+MCI) and control (CN) groups separately.
Then we discarded white matter, ventricular and cerebellar regions, remaining with 82 regions. we averaged SUVr of each region from both hemispheres; among the 41 remained regions, we averaged together the ones belonging to the same lobe/the same subcortical region. We selected 11 macro-regions, i.e. frontal, temporal, parietal, cingulate, thalamus, caudate, putamen, pallidum, hippocampus, amygdala, accumbens, and averaged together all the values of ROIs mapped to the same macro-region.
The macro--region definition was done to reduce computational expenses and aid interpretability of the resulting MP parameters.
Demographic and clinical details are shown in Table 3.
We split the data set in two parts: the D1 data set contains all the longitudinal data for each subject up to the second-to-last time points. The remaining time--points were included in a second data set D2. Subjects with one measurement only were included in D1. Data set D1 includes 1651 longitudinal data of 1091 subjects; D2 contains 731 cross-sectional data. We run the models on D1, estimating MP dynamics, GP parameters and individual time--shifts, and used D2 to validate model predictions.

\begin{table}[!htb]
    \centering
    \begin{tabular}{c| c c c}
        \hline
         Group & CN & MCI & AD  \\
        \hline
         N (female) & 369 (194)&526 (229)&195 (79)\\
         age (std) & 73.5 (6.0) & 72.0 (7.5) & 73.8 (7.7)\\
         years education (std) & 16.5 (2.6) & 16.1 (2.7) & 15.8 (2.7)\\
         APOE4 + & 102 & 244 & 129\\
         ADAS13 (std)& 9.1 (4.6) & 14.9 (6.9) & 31.3 (9.6)\\
         FAQ (std)& 0.3 (1.3) & 2.7 (4.0) & 14.1 (7.1)\\
         MMSE(std) & 29.0 (1.2) & 28.0 (1.8) & 22.6 (3.2)\\
         RAVLT learning (std)& 5.9 (2.4) & 4.7 (2.6) & 1.9 (1.8)\\
         \hline
    \end{tabular}
    \caption{Baseline socio demographic and clinical information the study cohort (1090 subjects for 2380 time--points). ADAS13: Alzheimer's Disease Assessment Scale-cognitive subscale, 13 items; FAQ: Functional Assessment Questionnaire; RAVLT learning: Rey Auditory Verbal Learning Test, learning item.}
    \label{tab:demographics}
\end{table}
\subsubsection{HCP data.}
Data used in the preparation of this work were obtained from the MGH-USC Human Connectome Project database. We collected 3D T1w and DTI of 24 age and gender--matched subjects. The pipeline for structural connectome generation is described in \cite{oxtoby2017datadriven}. We averaged the 24 connectomes together and obtained an average young, healthy connectome. Finally, we averaged together the regions belonging to the same lobe or subcortical area (to obtain 11 macro--region) and we set to 0 all the weights below the average weights across nodes, and to 1 the weight above. This last step was performed in order to remove the weak connections.

\subsection{Estimated long term dynamics.}
We analyzed the AV45--PET data with two different models of MP kinetics: the ACP model of equation (\ref{eq:mainmatr}) - which has non--linear and time--varying dynamics, and a full diffusive model. The diffusion dynamics were prescribed by the system 
$\boldsymbol{\dot{c}}(t) = B \boldsymbol{c}(t)$, where the coefficients $b_{ij}$ of $B$ are estimated (along with individual time parameters) with our framework.
Figure 3A) shows the long term trajectories estimated with both models, for four regions. Full results are in the Supplementary Material. Figure 3B) shows the time associated to each regional trajectory at which maximal separation between ``Cognitively Normal'' and ``Alzheimer's disease'' subjects was measured. The time distribution is inferred from the trajectory distribution associated to each region. The dynamics and orderings of ACP and diffusion provide plausible description of the pathological evolution of amyloid deposition, compatible with previous findings in histo pathological and imaging studies in AD \cite{thal2002phases,irvine2008protein}.
\begin{figure}[htb!]
    \centering
    \begin{tabular}[t]{c}
    {\includegraphics[width=4in]{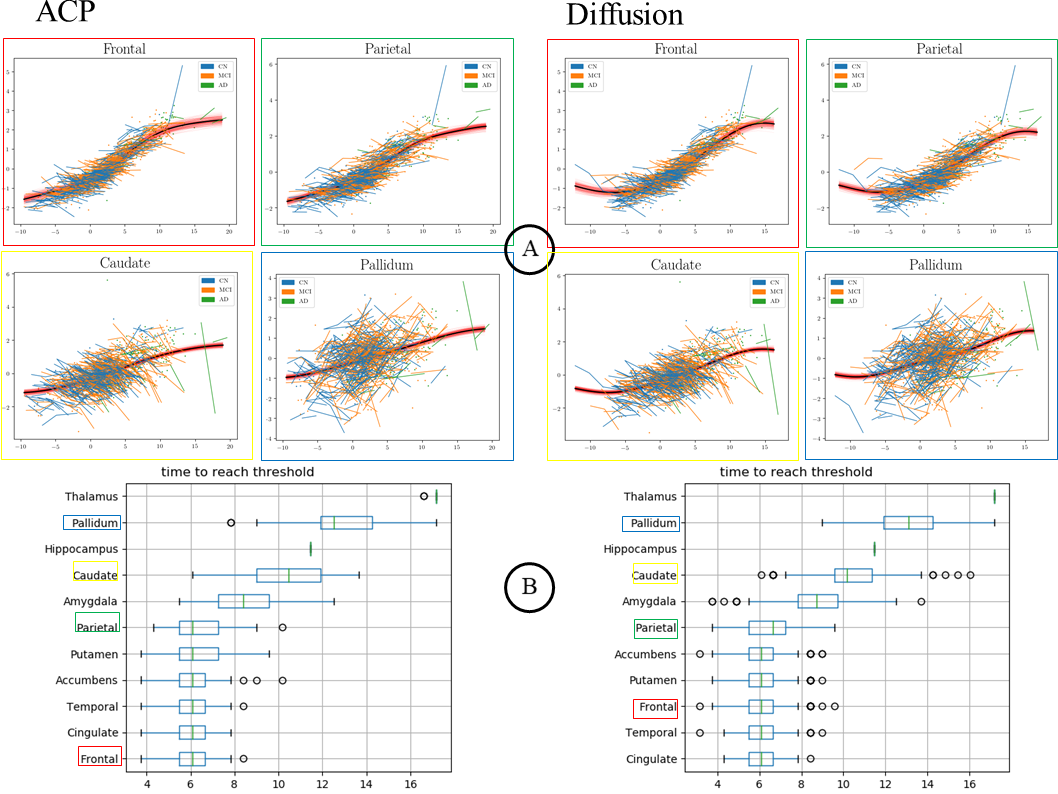}}\\[-0.2cm]
    {\includegraphics[width=4in]{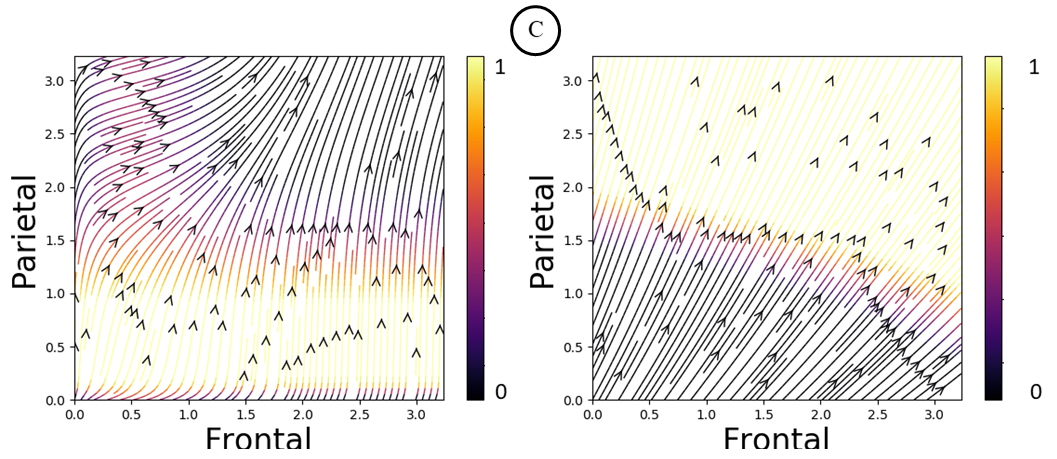}}
    \end{tabular}
    \caption{A): estimated long term trajectories and individual short term measurements for 4 regions of interest: frontal lobe, parietal lobe, caudate and pallidum, for the two models. B): ordering derived from the trajectories. Regions visualized in A) are highlighted. C): streamlines of the 2D fields in the $\{$Frontal, Parietal$\}$ plane.}
    \label{fig:my_label}
\end{figure}
\subsection{Predictions performances of the models.}
Figure 3C) shows the estimated vector fields for the relative dynamics of the frontal and parietal lobes. This vector field is obtained by integrating the dynamical system estimated for respectively ACP and diffusion models. Therefore it does not correspond to extrapolation of the curves in Figure 3A).
Here the other biomarkers are set constant to their mean values. We can appreciate the non--linear dynamics of the ACP model, as well as the linear dynamics of the diffusive model.
The resulting vector fields provide a tool for interpreting and comparing mechanistic hypotheses.
Indeed, Figure 3C) shows that the ACP model estimates an initial fast propagation, which slows down with time. The opposite behaviour is observed by analyzing the dynamics of the diffusion model, with an acceleration in the propagation along with the progression. This behaviour is unlikely to reproduce real--case scenarios, where amyloid aggregation eventually slows down and does not accumulates indefinitely. This result points to the higher biological plausibility of the proposed ACP model. 
For each subject in D1 with follow-up measurements in D2, we computed the streamline associated with their individual dynamics (in the whole 11-D space), and estimated the values of each biomarker at the corresponding follow-up time. We computed the RMSE for each estimate, and bootstrapped over the MP dynamic parameters 200 times, obtaining RMSE distributions (Table 4).
\begin{table}[htb!]
    \centering
    \hspace{-2cm}
    \begin{tabular}{c|c|c|c|c}
         &frontal& temporal& parietal& cingulate\\
         \hline
          ACP &0.21(0.16)&0.18(0.14)&0.20(0.16)&0.20(0.15)\\
          diffusion&0.25(0.18)&0.25(0.17)&0.22(0.19)&0.24(0.18)\\
    \end{tabular}
    \begin{tabular}{c|c|c|c|c|c|c|c}
    & thalamus& caudate& putamen& pallidum& hippo& amygdala& accumbens\\
    \hline
      ACP&0.12(0.09)&0.16(0.12)&0.16(0.13)&0.13(0.10)&0.12(0.09)&0.12(0.10)&0.21(0.15)\\
    diffusion&0.11(0.08)&0.17(0.13)&0.17(0.13)&0.17(0.13)&0.12(0.09)&0.11(0.09)&0.24(0.19) \\
    \end{tabular}
    \caption{RMSE (mean, sd) for the ACP and the diffusion models estimates. The ACP model generally provides predictions closer to the observed follow-up values.}
    \label{tab:RMSE_ACPvsDIFF}
\end{table}
\subsection{Misfolded proteins propagation pathways.}
Figure 4 shows the connectomes where the edges' colors are set to be proportional to the values of the estimated MP parameters for the ACP model (plot on the left hemisphere), or to the values of the propagation parameters for the diffusive model (plot on the right hemisphere). The parameters were normalized to $[0,1]$ to aid comparison. The paths appear to be different for the two models and the ACP model seems to better describe the frontal--posterior pathway known to characterize amyloid deposition in AD \cite{thal2002phases,irvine2008protein}.

\begin{figure}[htb!]\label{fig:brains}
    \centering
    \includegraphics[width=2.4in]{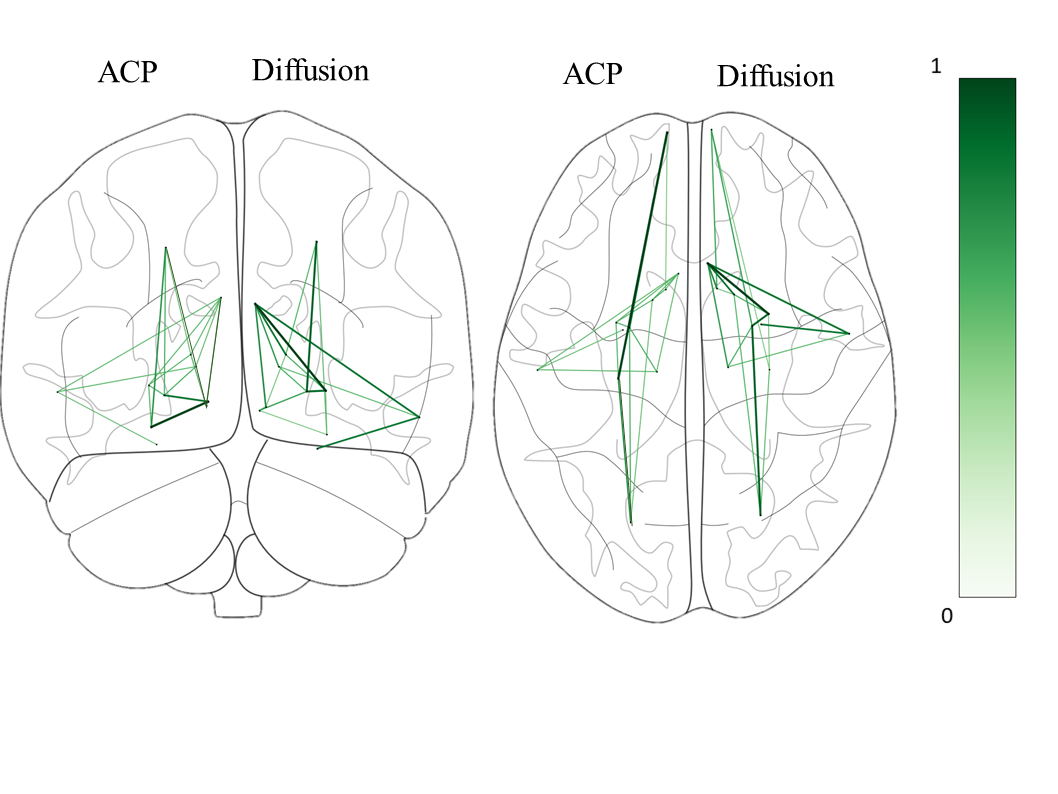}
    \caption{Coronal and axial views of connectomes with edges' colors proportional to the values of the estimated propagation parameters for either the ACP model (left hemisphere) and the diffusion model (right hemisphere).}
\end{figure}
\vspace{-1cm}

\section{Discussion}
We presented a spatio--temporal model of MP dynamics over brain networks. The model is based on the joint estimation of long term MP dynamics and time reparametrization of individuals observations, and is expressed within a GP regression setting, where constraints on the MP dynamics are imposed through non--linear dynamical systems, which account for accumulation, clearance and propagation of MP. 
Experiments  on  simulated  data  show  that  our  model  accurately  recovers  prescribed  rates along  graph  dynamics  and  precisely reconstructs the underlying progression.  When applied to AV45-PET brain imaging data our model allows the bio-mechanical interpretation of amyloid deposition in Alzheimer's disease, leading to plausible simulations of MP propagation, and achieving accurate predictions of individual MP deposition in unseen data. 

The method has some limitations: first of all, structural connectome estimation using tractography is known to be prone to false positive and negative connections. Nevertheless, here we take an average connectome over multiple young and healthy subjects, which we believe provides a reasonable anatomical reference. Another limitation of the model is that it assumes that all subjects follow the same disease progression pattern, which might not be the case in heterogeneous data sets such as ADNI. 

The ideas we propose here extend to a much larger range of diseases and alternative models of propagation, such as propagation via functional networks \cite{zhou2012predicting,cauda2018brain}, or different kind of tractography to represent intra-- and extra--axonal propagation \cite{oxtoby2017datadriven,iturria2018multifactorial}. 

\section*{Acknowledgments}
SG acknowledges financial support from the French government managed by L'Agence Nationale de la Recherche under Investissements d'Avenir UCA JEDI (ANR-15-IDEX-01) through the project ``AtroProDem: A data-driven model of mechanistic brain Atrophy Propagation in Dementia''.

This work is supported by the Inria Sophia Antipolis - M\'editerran\'ee, ``NEF'' computation cluster.

Data collection and sharing for this project was funded by the Alzheimer’s Disease Neuroimaging Initiative (ADNI) (National Institutes of Health Grant U01 AG024904). ADNI is funded by the National Institute on Aging, the National Institute of Biomedical Imaging and Bioengineering, and through generous contributions from the following: AbbVie, Alzheimer’s Association; Alzheimer’s Drug Discovery Foundation; Araclon Biotech; BioClinica, Inc.; Biogen; Bristol- Myers Squibb Company; CereSpir, Inc.; Eisai Inc.; Elan Pharmaceuticals, Inc.; Eli Lilly and Company; EuroImmun; F. Hoffmann-La Roche Ltd and its affiliated company Genentech, Inc.; Fujirebio; GE Healthcare; IXICO Ltd.; Janssen Alzheimer Immunotherapy Research and Development, LLC.; Johnson and Johnson, Pharmaceutical Research and Development LLC.; Lumosity; Lundbeck; Merck and Co., Inc.; Meso Scale Diagnostics, LLC.; NeuroRx Research; Neurotrack Technologies; Novartis Pharmaceuticals Corporation; Pfizer Inc.; Piramal Imaging; Servier; Takeda Pharmaceutical Company; and Transition Therapeutics. The Canadian Institutes of Health Research is providing funds to support ADNI clinical sites in Canada. Private sector contributions are facilitated by the Foundation for the National Institutes of Health (www.fnih.org). The grantee organization is the Northern California Institute for Research and Education, and the study is coordinated by the Alzheimer’s Disease Cooperative Study at the University of California, San Diego. ADNI data are disseminated by the Laboratory for NeuroImaging at the University of Southern California.

The HCP project (Principal Investigators: Bruce Rosen, M.D., Ph.D., Martinos Center at Massachusetts General Hospital; Arthur W. Toga, Ph.D., University of California, Los Angeles, Van J. Weeden, MD, Martinos Center at Massachusetts General Hospital) is supported by the National Institute of Dental and Cranio-facial Research (NIDCR), the National Institute of Mental Health (NIMH) and the National Institute of Neurological Disorders and Stroke (NINDS). Collectively, the HCP is the result of efforts of co-investigators from the University of California, Los Angeles, Martinos Center for Biomedical Imaging at Massachusetts General Hospital (MGH), Washington University, and the University of Minnesota.

\bibliographystyle{unsrtnat}

\begin{thebibliography}{5}

\bibitem{soto2012protein}
Soto, C. and Pritzkow, S., 2018. Protein misfolding, aggregation, and conformational strains in neurodegenerative diseases. Nat Neurosci, 21(10), pp.1332-1340.
\bibitem{junker2013self}
Jucker, M. and Walker, L.C., 2013. Self-propagation of pathogenic protein aggregates in neurodegenerative diseases. Nature, 501(7465), p.45.
\bibitem{carbonell2018mathematical}
Carbonell, F., Iturria-Medina, Y. and Evans, A.C., 2018. Mathematical modeling of protein misfolding mechanisms in neurological diseases: a historical overview. Frontiers in Neurology, 9, p.37.
\bibitem{raj2012network}
Raj, A., Kuceyeski, A. and Weiner, M., 2012. A network diffusion model of disease progression in dementia. Neuron, 73(6), pp.1204-1215.
\bibitem{oxtoby2017datadriven}
Oxtoby, N.P., Garbarino, S., Firth, N.C., Warren, J.D., Schott, J.M., Alexander, D.C. and Alzheimer’s Disease Neuroimaging Initiative, 2017. Data-Driven sequence of changes to anatomical Brain connectivity in sporadic Alzheimer’s Disease. Frontiers in neurology, 8, p.580.
\bibitem{zhou2012predicting}
Zhou, J., Gennatas, E.D., Kramer, J.H., Miller, B.L. and Seeley, W.W., 2012. Predicting regional neurodegeneration from the healthy brain functional connectome. Neuron, 73(6), pp.1216-1227.
\bibitem{iturria2018multifactorial}
Iturria-Medina, Y., Carbonell, F. M., Sotero, R. C., Chouinard-Decorte, F. and Evans, A. C., 2017 Multifactorial causal model of brain (dis)organization and therapeutic intervention: Application to Alzheimer’s disease. Neuroimage 152, pp.60-77.
\bibitem{cauda2018brain}
Cauda, F., Nani, A., Manuello, J., Premi, E., Palermo, S., Tatu, K., Duca, S., Fox, P.T. and Costa, T., 2018. Brain structural alterations are distributed following functional, anatomic and genetic connectivity. Brain, 141(11), pp.3211-3232.
\bibitem{young2014datadriven}
Young, A.L., Oxtoby, N.P., Daga, P., Cash, D.M., Fox, N.C., Ourselin, S., Schott, J.M. and Alexander, D.C., 2014. A data-driven model of biomarker changes in sporadic Alzheimer's disease. Brain, 137(9), pp.2564-2577.
\bibitem{lorenzi2017probabilistic}
Lorenzi, M., Filippone, M., Frisoni, G.B., Alexander, D.C., Ourselin, S. and Alzheimer's Disease Neuroimaging Initiative, 2017. Probabilistic disease progression modeling to characterize diagnostic uncertainty: application to staging and prediction in Alzheimer's disease. Neuroimage.
\bibitem{schiratti2017bayesian}
Schiratti, J.B., Allassonnière, S., Colliot, O. and Durrleman, S., 2017. A Bayesian mixed-effects model to learn trajectories of changes from repeated manifold-valued observations. The Journal of Machine Learning Research, 18(1), pp.4840-4872.
\bibitem{donohue2014estimating}
Donohue, M.C., Jacqmin-Gadda, H., Le Goff, M., Thomas, R.G., Raman, R., Gamst, A.C., Beckett, L.A., Jack Jr, C.R., Weiner, M.W., Dartigues, J.F. and Aisen, P.S., 2014. Estimating long-term multivariate progression from short-term data. Alzheimer's and Dementia, 10(5), pp.S400-S410.
\bibitem{lorenzi2018constraining}
Lorenzi, M. and Filippone, M., 2018. Constraining the Dynamics of Deep Probabilistic Models. Proceedings of the 35th International Conference on Machine Learning, 80, pp.3233-3242.
\bibitem{cutajar2017random}
Cutajar, K., Bonilla, E. V., Michiardi, P., and Filippone, M., 2017.  Random feature expansions for deep Gaussian processes. Proceedings of the 34th International Conference on Machine Learning, 70, pp.884-893.
\bibitem{thal2002phases}
Thal, D.R., Rub, U., Orantes, M. and Braak, H., 2002. Phases of Aβ-deposition in the human brain and its relevance for the development of AD. Neurology, 58(12), pp.1791-1800.
\bibitem{irvine2008protein}
Irvine, G.B., El-Agnaf, O.M., Shankar, G.M. and Walsh, D.M., 2008. Protein aggregation in the brain: the molecular basis for Alzheimer’s and Parkinson’s diseases. Molecular medicine, 14(7-8), pp.451-464.








\end{thebibliography}

\newpage
\section*{Supplementary Material}
\subsection*{Details on the implementation}
 
The optimization is performed with stochastic gradient descent with adaptive moment estimation (Adam), through the alternate optimization of i) the approximated posterior over $\boldsymbol{W}$, the likelihood parameters for the GP and for the random effect, ii) the individual time--shifts parameters, and iii) the approximated posterior over the MP parameters and the likelihood parameters of the constraints. Also, for the MP parameters scheme, we split optimization over $\bar{k}_{ij}$ and $\bar{k}_t$, and over $\eta_i$ and $\gamma_i$. This is done to reduce the variability of their gradients, as such parameters have values in different ranges. 
Further, we first run the optimization scheme for estimating values of the two brain--level parameters $\eta$ and $\gamma$, and then estimate $\eta_i$ and $\gamma_i$ using as starting points the estimated values of $\eta$ and $\gamma$ and lowering the learning rate.

\subsection*{Simulation on single--subject}

Figure 5 shows examples of long term trajectories that is possible to obtain by varying the MP parameters of the ACP model, on a specified time interval, for a two--region problem.
\begin{figure}[ht]\label{fig:multiple2D}
    \centering
    \begin{tabular}{ccc}
        \includegraphics[width = 1.5in]{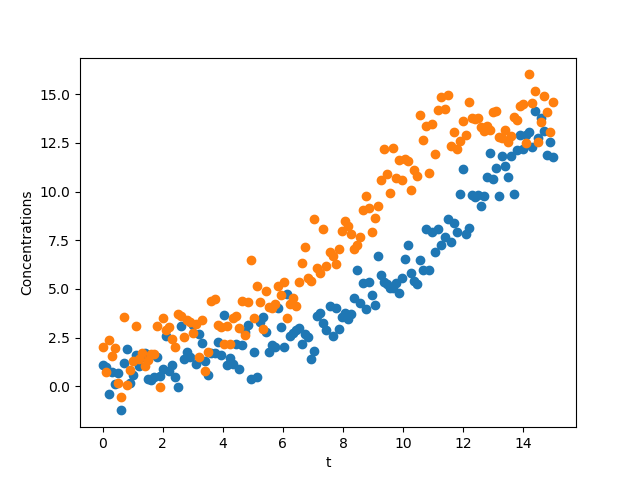} &
        \includegraphics[width = 1.5in]{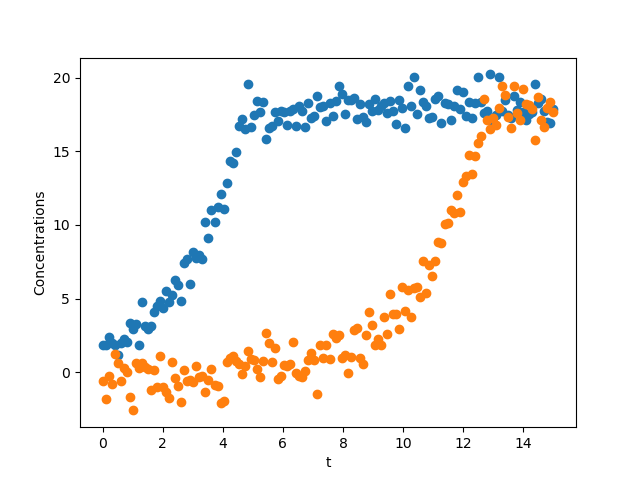} &
        \includegraphics[width = 1.5in]{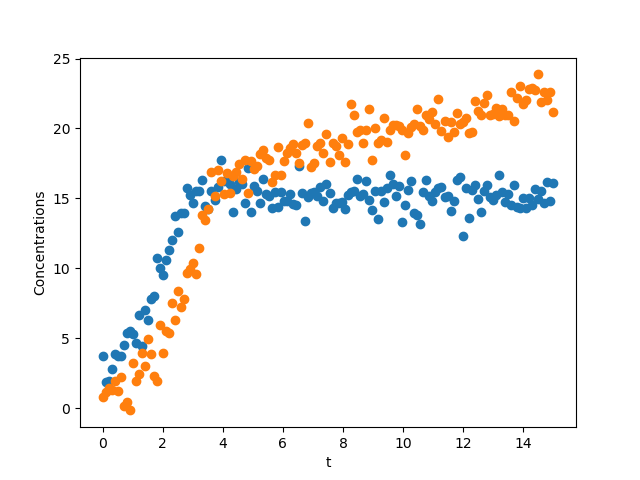}\\
        \includegraphics[width = 1.5in]{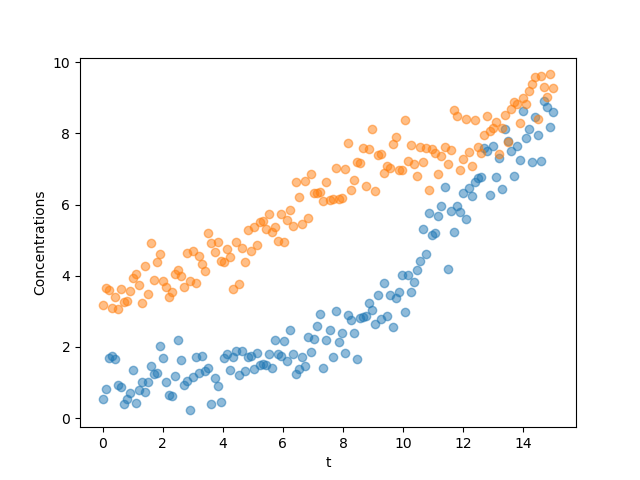} &
        \includegraphics[width = 1.5in]{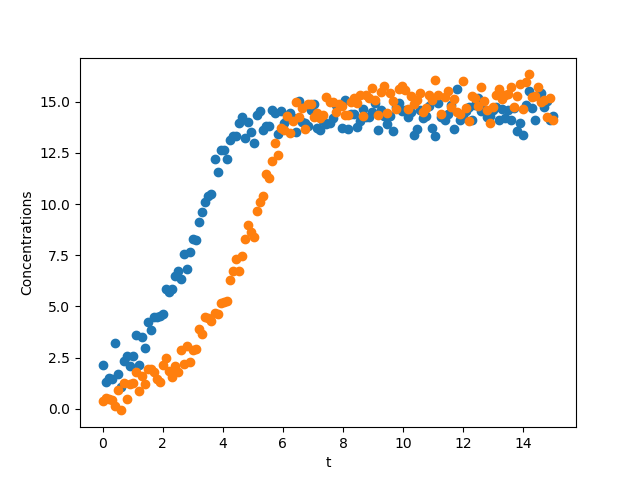} &
        \includegraphics[width = 1.5in]{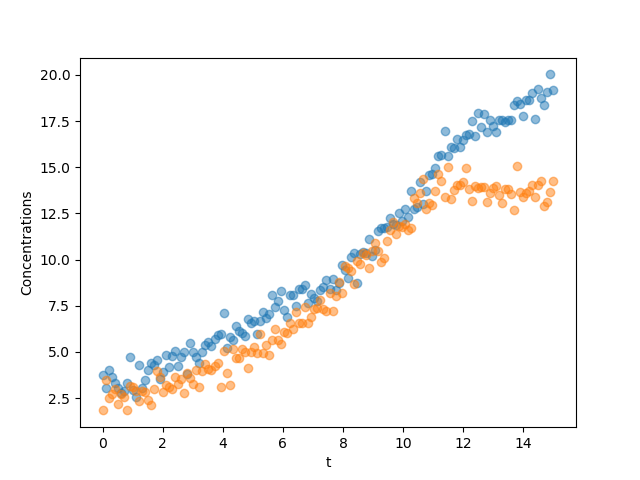}
    \end{tabular}
    \caption{Examples of (noisy) long--term trajectories obtained by varying the MP parameters and the noise, on a time interval [0,15].}
\end{figure}
Figure 6A) and 6B) show the results of the fit for the example data in Figure 2 top left in term of the concentration evolution over time of the MP dynamics and of the MP parameter.
\begin{figure}[!htb]\label{fig:results2D}
    \centering
    \begin{tabular}[t]{c}
        \includegraphics[width=5in]{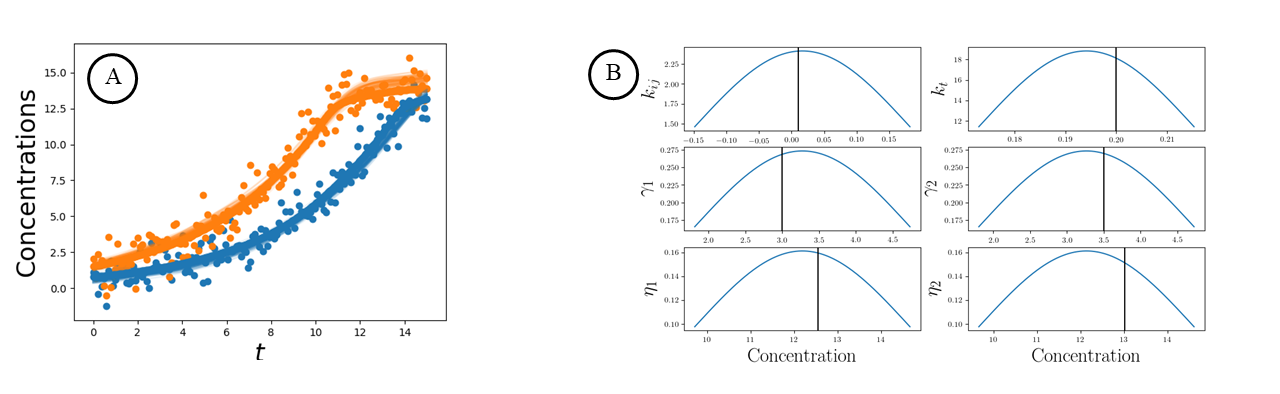}
    \end{tabular}
    \caption{Results for example data in Figure 5(a). A) data as circles and estimated dynamics as lines. B) estimated distribution of $\kij$, $k_t$, $\gamma$ and $\eta$ (2-$\sigma$ shown here), and their ground truth values as vertical black lines.}
\end{figure}

In order to evaluate the accuracy of the model in reconstructing the parameters for increasing number of regions, we generated other five two-dimensional problems varying the MP parameters and the noise (as in Figure 5), and repeated the procedure for increasing number of regions: 3, 5, 11 and 42. Table 5 shows the distributions of the rooted mean square errors (RMSEs) for each MP parameter and for the GP parameters (estimated VS ground truth parameters), at increasing number of regions. We note that the RMSEs are always close to zero, with the smallest error consistently on $k_t$ and the biggest on $\eta$ or $\gamma$. 

\begin{table}[htb!]
    \centering
    \begin{tabular}{cccccc}
         &N=2 & N=3 &  N=11 & N=42 \\
        \hline
        \hline
        $\bar{k}_{ij}$ & 1.4\% & 1.4\%& 4.4\%&4.4\%\\
        \hline
        $\bar{k}_t$ & 0.3\%& 0.3\% &0.1\%&0.1\%\\
        \hline
        $\gamma$ & 11.3\% & 11.5\%&18.0\%&23.2\%\\
        \hline
        $\eta$ & 14.7\% & 16.9\%&24.2\%&23.0\%\\
        \hline
        Data fit & 7.4\%& 8.4\%&8.5\%&10.2\%\\
    \end{tabular}
    \caption{RMSE results for dynamical parameters and GP fit estimates for the ACP model. The errors are expressed, in percentage, relatively to the ground truth parameters.}
    \label{tab:RMSE_single subject}
\end{table}
\subsection*{Full results for the real data scenario}
Figure 7 shows the estimated long term trajectories for all the 11 macro-regions, in the time interval [-10, 15], for both the models. 
\begin{figure}[htb!]
    \centering
    {\includegraphics[width=5in]{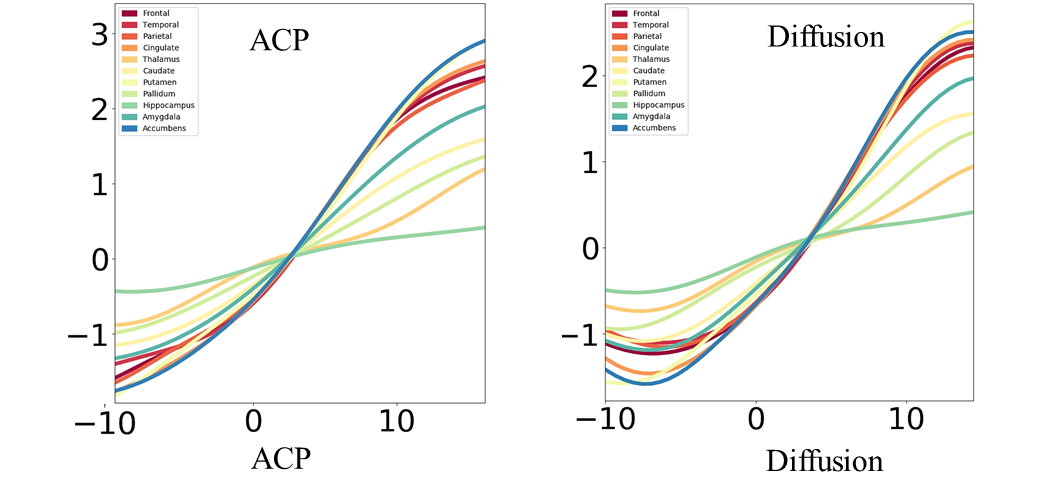}}
    \caption{estimated long term trajectories for the ACP (left) and the diffusion (right) models.}
    \label{fig:all_traj}
\end{figure}

\end{document}